\documentclass[structabstract]{aa}

\usepackage{url}
\usepackage[breaklinks=true]{hyperref}
\usepackage{twoopt}
\usepackage[english]{babel}          

\usepackage{natbib}
\bibpunct{(}{)}{;}{a}{}{,} 

\usepackage[utf8]{inputenc}
\usepackage[normalem]{ulem}
\usepackage{siunitx}
\usepackage{amsmath, amssymb}

\usepackage{bigstrut}
\usepackage[varg]{txfonts} 
\usepackage{graphicx}
\usepackage{threeparttable}

\def \kms {km$\,$s$^{-1}$}

\def \deg     {\text{$^{\circ}$}}
\def \arcmin      {\text{$^\prime$}}
\def \arcsec      {\text{$^{\prime\prime}$}}

\def \mjybeam     {mJy\,beam$^{-1}$}
\def \mujybeam    {$\mathrm{\mu}$Jy\,beam$^{-1}$}

\newcommand{\beam}[2]{{#1}\arcsec$\times${#2}\arcsec}

\newcommand{\Msun}{\text{$\rm M_\odot$}}
\newcommand{\Rsun}{\text{$\rm R_\odot$}}
\newcommand{\hdstar}{\object{HD~80606}}
\newcommand{\planet}{\object{HD~80606b}}

\begin{document}

\title{Radio observations of HD~80606 near planetary periastron}
\subtitle{II.~LOFAR low band antenna observations at~30--78~MHz}
\titlerunning{Radio Observations of HD~80606 Near Planetary Periastron}

\author{F.~de~Gasperin\inst{1,2}
\and T.~J.~W.~Lazio\inst{3}
\and M. Knapp\inst{4}} 

\institute{Hamburger Sternwarte, Universit{\"a}t Hamburg, Gojenbergsweg~112, 21029, Hamburg, Germany \and
INAF - Istituto di Radioastronomia, via P. Gobetti 101, Bologna, Italy \and
Jet Propulsion Laboratory, California Institute of Technology, 4800 Oak Grove Dr., M/S~67-201, Pasadena, CA 91109 \and
MIT Haystack Observatory, 99 Millstone Rd., Westford, MA, 01886}

\date{Received 2020 / Accepted ...}

\abstract
{All the giant planets in the Solar System generate radio emission via electron cyclotron maser instability, giving rise most notably to \object{Jupiter}'s decametric emissions.  An interaction with the solar wind is at least partially responsible for all of these Solar System electron cyclotron masers. \planet{} is a giant planet with a highly eccentric orbit, leading to predictions that its radio emission may be enhanced substantially near periastron.}
{This paper reports observations with the Low Frequency Array (LOFAR) of \planet{} near its periastron in an effort to detect radio emissions generated by an electron cyclotron maser instability in the planet's magnetosphere.}
{The reported observations are at frequencies between~30~MHz and~78~MHz, and they are distinguished from most previous radio observations of extrasolar planets by two factors: (i)~They are at frequencies near~50~MHz, much closer to the frequencies at which Jupiter emits ($\nu < 40\,\mathrm{MHz}$) and lower than most previously reported observations of extrasolar planets; and~(ii)~sensitivities of approximately a few millijanskys have been achieved, an order of magnitude or more below nearly all previous extrasolar planet observations below~100~MHz.}
{We do not detect any radio emissions from \planet{} and use these observations to place new constraints on its radio luminosity.  We also revisit whether the observations were conducted at a time when \planet{} was super-Alfv{\'e}nic relative to the host star's stellar wind, which experience from the Solar System illustrates is a state in which an electron cyclotron maser emission can be sustained in a planet's magnetic polar regions.}
{}


\maketitle

\section{Introduction}
\label{sec:intro}

Giant planets in the Solar System generate radio emission as a result, at least partially, of the interaction between the solar wind and the planet's magnetospheres. The solar wind impinging on a planet's magnetosphere {can generate} currents {in} the planet's magnetic polar regions, where an electron cyclotron maser is produced.  There is a rich literature on the relevant processes, based substantially on \textit{in situ} observations by spacecraft, and straightforward extrapolations predict that the same processes {could occur} in extrasolar planetary systems \citep[e.g.,][]{1986ApJ...309L..59W,1997pre4.conf..101Z,1999JGR...10414025F,2000ApJ...545.1058B,2001Ap&SS.277..293Z,2007P&SS...55..598Z}.  
The detection of magnetospherically generated radio emission in these systems would be the first direct detection of magnetic fields in extrasolar planets.

Radio emissions from extrasolar planets have been extensively searched for at frequencies ranging from 25~MHz to 1400~MHz, with most of these observations having focused on ''hot Jupiters,'' massive planets in orbits with small semi-major axes and low eccentricities; both \cite{2015aska.confE.120Z} and 
\cite{2016pmf..rept.....L} contain extensive summaries of observational results.
Sensitivities ($1\sigma$) obtained in past searches for extrasolar planetary magnetospheric emission have ranged from roughly 1000~mJy to below 1~mJy, with generally better sensitivities obtained at the higher frequencies. Typically, these sensitivities have not been sufficient to challenge the predicted levels, though some multi-epoch 74~MHz observations of $\tau$ Boo~b \citep{2007ApJ...668.1182L} do indicate that its emission is not consistent with the most optimistic predictions, if it does emit at that frequency.
Recently, \citet{2020NatAs...4..577V} reported low radio frequency emission from the quiescent late-type star GJ~1151.  The characteristics of this emission are consistent with it being generated by a sub-Alfv{\'e}nic flow into the star's magnetic polar regions, potentially from an orbiting terrestrial-mass planet.  While exciting (and these observations would be sensitive to similar kinds of emission), such emissions do not clearly lead to constraints on the properties of the planet itself.

Moreover, existing estimates regarding the magnetic fields of hot Jupiters have considerable uncertainties, potentially at the level of more than an order of magnitude \citep[e.g.,][]{2010ApJ...712.1277F, 2010ApJ...709..670E,Kao_2016,2019NatAs...3.1128C,Yadav_2017}.
If hot Jupiter magnetic field strengths are comparable to, or weaker than, that of Jupiter, the range of frequencies over which their electron cyclotron masers would operate are unlikely to extend much above the range for Jupiter ($\la 40$~MHz). Consequently, many of the existing observations may be at {frequencies} that are too high to detect extrasolar electron cyclotron maser emission \citep{2012MNRAS.423.3285V}.

The rationale behind the general focus on hot Jupiters has been twofold: If an extrasolar planet is close to its host star, the stellar wind should be more intense, with a concomitant increase in the strength of the planetary radio emission \citep{1999JGR...10414025F, 2001Ap&SS.277..293Z, 2007P&SS...55..598Z, 2018A&A...618A..84Z}.  Secondarily, planets close to their host stars produce the largest radial velocity signature, and, until the \textit{Kepler} mission, the vast majority of extrasolar planets had been found with the radial velocity method. 

An alternative line of investigation is to focus on highly eccentric extrasolar planets; the dramatic increases in stellar wind pressures that these planets experience near their periastrons should result in substantial enhancements in their radio luminosities \citep{1981GeoRL...8.1087G, 1983JGR....88.8999D, 1984JGR....89.6819D}.
A prototypical example is the giant planet \planet{} ($M=3.94\pm0.11$~M$_J$, $d=66.6$~pc), which has a 111~day orbital period with one of the highest known orbital eccentricities \citep[$e = 0.9336 \pm 0.0002$;][]{2009MNRAS.396L..16F}. Physical modeling of fluid bodies (the star and planet) in an eccentric orbit predicts that the planet should be driven into a state of pseudo-synchronized rotation \citep{1981A&A....99..126H}, with a rotational period of 39.9~hr.

With a mass and a rotation period comparable to those of Jupiter, \planet{} could reasonably be expected to emit at or above the frequencies at which Jupiter does ($\sim 40$~MHz). \cite{Lazio2010} estimated that its emissions could extend as high as 55~MHz, and potentially to 90~MHz, when  {uncertainties on its mass and radius (i.e., bulk density) are considered}. Comparing the sizes of the orbits of Jupiter and \planet{}, and using what is known about the strength of the stellar wind of \planet{}, \cite{Lazio2010} estimated the range of planetary luminosities during the course of its orbit. Near periastron, \planet{} could reasonably be expected to be 3000 times as luminous as Jupiter, reaching a flux density of $\sim 1$~mJy when observed from Earth.

This paper reports observations of \planet{} close to its periastron at a central frequency of 54~MHz ($\lambda$5.45~m).
{This work improves upon that of \cite{Lazio2010} in two aspects.  First, it is both at a lower frequency, more comparable to that at which Jupiter emits, and it is more sensitive.  Second, these observations include analyses of both the total and the circularly polarized intensities.}
An initial analysis of some of these data, using different methods, was presented by \cite{k18}, {who obtained} an rms noise of 6~\mjybeam{}. In Sect. \ref{sec:observe}, we summarize the observations and data analyses, in Sect.\ \ref{sec:results} we present our results, and in Sect.\ \ref{sec:conclude} we discuss the results in context and present our conclusions.

\section{Observations and analysis}
\label{sec:observe}

Table~\ref{tab:log} presents the observing log. The observations consist of {five} epochs that occurred during Low Frequency Array (LOFAR) Cycle~4. The observing strategy was to obtain a reference observation, near apoastron, to serve as a baseline observation. There are two observations prior to periastron, occurring approximately 48~hr and 18~hr prior, and two observations subsequent to periastron, occurring approximately 48~hr and 18~hr after.  Observations {following} periastron were conducted to allow for the possibility that an increase in planetary {radio} luminosity lags the increase in incident stellar wind.  The combined uncertainties in the orbital periods and times of periastron typically amount to less than 0.1~days (about~2~hr).

\begin{table*}
\centering
\caption{\planet{} observing log.}\label{tab:log}
\begin{threeparttable}
\begin{tabular}{llcc}
Epoch & Orbital position & Usable Hours & Elevation\bigstrut[t]\\
\hline\hline
2015 August~30  & Near apastron            &  6 & 46\deg -- 87\deg \\
\\
2015 October~24 & 48~hours pre-periastron  &  0\tnote{a} & \ldots \\ 
2015 October~25 & 18~hours pre-periastron  &  4 & 30\deg -- 70\deg \\ 
\\
2015 October~27 & 18~hours post-periastron &  2 & 45\deg -- 53\deg \\ 
2015 October~30 & 48~hours post-periastron &  5 & 32\deg -- 64\deg \\
\hline
\end{tabular}
\begin{tablenotes}
 \item[a] No good data were obtained on~2015 October~24 due to a combination of low-elevation and ionospheric conditions.
\end{tablenotes}
\end{threeparttable}
\end{table*}

Observations at periastron were not conducted. While such observations would have the maximum incident stellar wind flux, there is the risk that the plasma frequency of the stellar wind could be larger than the planet's electron cyclotron frequency {\citep{2017MNRAS.469.3505W}}. If so, the planetary radio emission would not be able to escape from the system.

\subsection{Data acquisition}\label{sec:acquire}

All observations were acquired using LOFAR's Low Band Array \citep[\hbox{LBA},][]{VanHaarlem2013}, with the 30~MHz--75~MHz tuning used. {The use of this frequency range maximizes the sensitivity by excluding the top and bottom of the band where the dipole response is suppressed}.

Each LOFAR LBA station consists of 96 crossed dipoles;  however, {for Dutch stations (core and remote),} only 48 dipoles can be used simultaneously due to hardware limitations at the station level.  For these observations, the ``LBA Outer'' configuration was used, in which only the data from the outer 48 dipoles are recorded. The Outer configuration simplifies the calibration by reducing the primary beam size, and the dipoles cross-talk. The primary beam for this configuration varies from approximately 3\deg{} to~5\deg{} across the band; we adopted a fiducial value of approximately 4\deg{} (full width at half maximum).

We used 23 core stations and 14 remote stations. International stations were not included in order to maintain a reasonable data volume and because subarcsecond resolution was not crucial. The longest baseline was therefore approximately 100~km, providing a nominal resolution at mid-band (54 MHz) of 12\arcsec.

The observations were conducted in multi-beam mode, with one beam continuously pointing at a calibrator (\object{3C~196}) and one beam continuously pointing at the target (\planet).
Data were acquired in~244 subbands, of width 0.195~MHz, with 64 frequency channels per subband (3~kHz per frequency channel). During correlation, an integration time of 1~s and a frequency resolution of~3~kHz (i.e., the native resolution of the subband frequency channels) were used. After Radio Frequency Interference (RFI) detection, data were averaged down to 4~s and four frequency channels per subband, each with a bandwidth of 48~kHz.

\subsection{Data reduction}\label{sec:reduce}

The processing of the calibrator visibility data generally followed the procedures described previously by \cite{deGasperin2019}. The calibrator visibilities were used to isolate a number of systematic effects that were then applied to the target data. These effects are the amplitude bandpass, the stations' clocks drifts, and the polarization delay between the X and Y dipoles. Together with these effects, an analytic estimation of the dipole beam effect on both amplitudes and phases was applied to the target data \citep{VanHaarlem2013}.

The self-calibration starts by obtaining an initial sky model from available surveys: TGSS \citep{Intema2017}, NVSS \citep{Condon1998}, WENSS \citep{Rengelink1997}, and VLSS \citep{Lane2014}. Using sources from these surveys, we estimated their spectral indices, where possible up to the second order, and extrapolated them to the LBA frequency range. An average total electron content (TEC) above the target field was estimated for each antenna and time step.  Uncertainties in the estimated \hbox{TEC}, and the resulting phase and delay uncertainties, are the strongest systematic effect still present in the data at this stage of the calibration.  Once corrected, we can estimate the Faraday rotation and second-order beam  effects.  Sources in the first side lobe were imaged and subtracted from the data before starting a direction-dependent calibration.

The final direction-dependent calibration followed \cite{2020A&A...642A..85D}. Here, the best available model was divided into "direction-dependent calibrators" (DDcals), a compact group of sources on the sky, each of which accounts for at least 2~Jy of flux density. In this case, we used seven DDcals. For each DDcal, a direction-dependent solver (Offringa et al. in prep.) present in DP3\footnote{\url{https://github.com/lofar-astron/DP3}} estimated the direction-dependent component of the TEC along the direction connecting each DDcal to each station. Finally, the field-of-view was imaged in facets with wsclean \citep{Offringa2014}, applying the correction estimated from the closest DDcal.
The combined observations produced the deepest image ever made at ultra-low frequencies, with sensitivities of~800~\mujybeam{} in the Stokes~I parameter and~700~\mujybeam{} in the Stokes~V parameter.

\section{Results}
\label{sec:results}

After removing the best estimation of systematic effects for each epoch, we imaged each independently, combining all data within an epoch in order to improve the signal-to-noise ratio. For each epoch, we imaged the entire band (30~MHz--78~MHz) as well as only the lower half (30~MHz--54~MHz).  The motivation for imaging only the lower half of the band is that Jupiter's radio emission truncates sharply above its cutoff frequency ($\approx 40$~MHz); combining all of the data across the entire frequency band risks decreasing the signal-to-noise ratio for any \planet\ radio emission if its cutoff frequency occurs in the lower half of the band. Images were produced in both Stokes~I and~V parameters as the electron cyclotron maser instability is highly circularly polarized \citep[e.g.,][]{2006A&ARv..13..229T,2011PhPl...18e6501V}.

\begin{figure*}
  \centering
  \includegraphics[width=\textwidth]{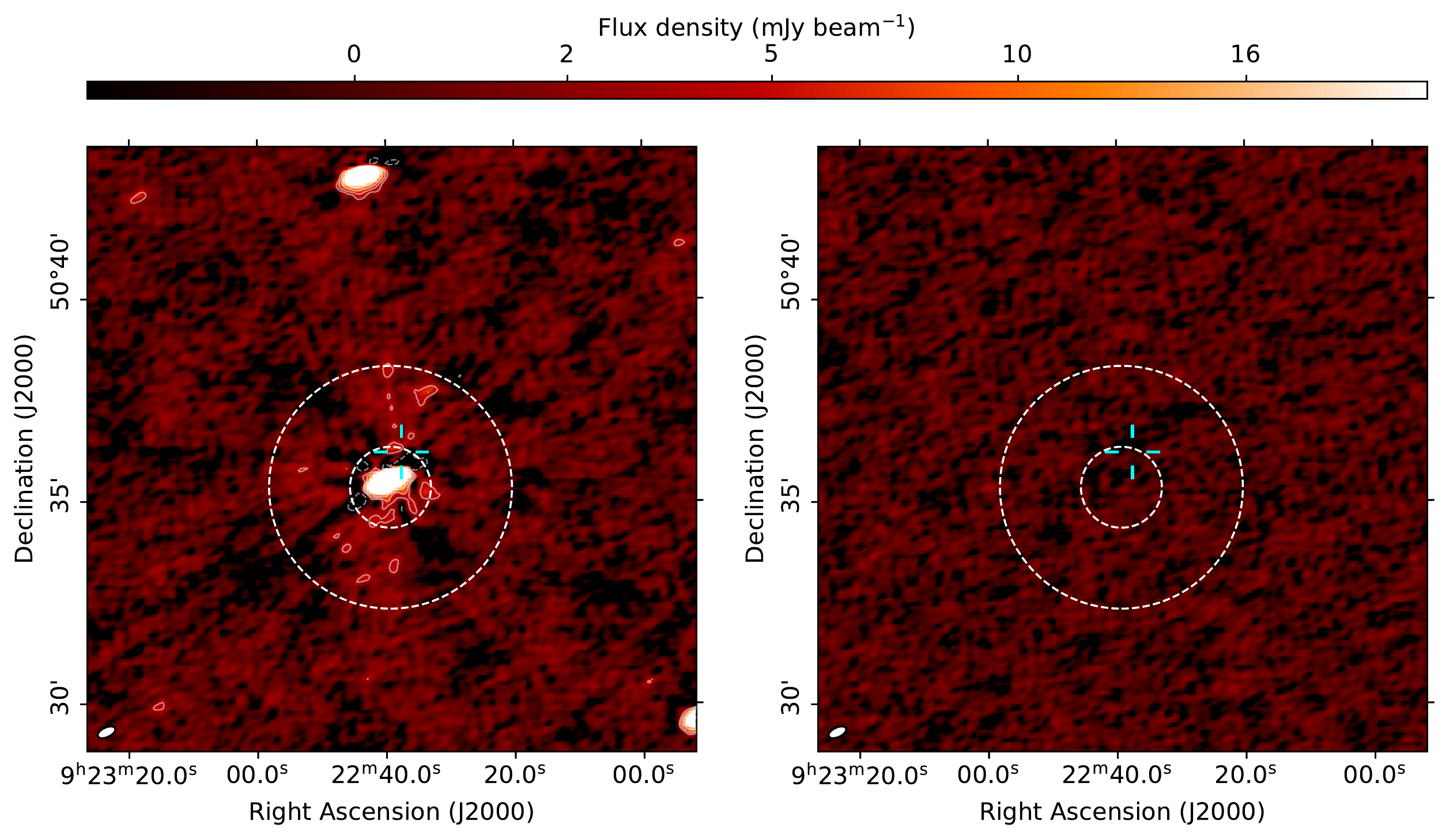}\hfil%
  \caption{Images of the immediate surroundings of \hdstar{}, for the full band (30~MHz--78~MHz). The location of the star is at the center of the cross-hair. The annulus traced by the dashed line is the region used to estimate the local rms noise. Contours start at $3\sigma$ with $\sigma=800$~\mujybeam. The beam, \beam{28}{14}, is shown in the bottom-left corner of the image. The source just to the south of the star's location is \object{FIRST~J092239.6$+$503529}. \textit{Left}: Stokes~I image. \textit{Right}: Stokes~V image, corresponding to the Stokes~I image shown in the left-hand panel.}
  \label{fig:reference}
\end{figure*}

At no epoch do we find evidence for radio emission from \planet{}. Table~\ref{tab:limits} summarizes the parameters of the images produced and the upper limits on the radio luminosity of the planet for each epoch, estimated using the full bandwidth (central frequency 54 MHz) and for the bottom half (central frequency 42 MHz). 
Variations in the image noise levels from epoch to epoch result from observing them at different elevations and from the changing of ionospheric conditions from day to day.  At each epoch we estimated a ``global'' rms noise level, using all non-source pixels in the image, and a ``local'' rms noise level, determined from an annulus with inner radius 1\arcmin{} and outer radius 3\arcmin{} centered around the nearby radio source \object{FIRST~J092239.6$+$503529} (see Fig.~\ref{fig:reference}). Figure~\ref{fig:reference} presents the Stokes~I and~V images of the \hdstar{} field for the reference epoch of~2015 August~30. For all images, the synthesized beam diameter was approximately \beam{30}{15}.

\begin{table*}
\centering
\begin{threeparttable}
\begin{tabular}{lccccc}
Epoch & Center & Stokes I & Stokes~I & Stokes~V & Luminosity \\
      & Frequency
    & global    & local    & global   & Limit ($3\sigma$)\\
    & (MHz)    & \multicolumn{3}{c}{(\mjybeam)} & (erg~s${}^{-1}$)\\
\hline\hline
2015 August~30 (off) & 54 & 1.2 & 1.6 & 1.0 & \ldots \\
                     & 42 & 2.0 & 2.6 & 1.5 & \ldots \\
\\
2015 October~24 (pre-periastron)\tnote{a}
    & \ldots & \ldots & \ldots & \ldots & \ldots \\
2015 October~25 (pre-periastron) 
    & 54 &  2.3 &  4.0 & 1.7 & $1.8 \times 10^{23}$ \\
    & 42 &  3.9 &  6.9 & 2.7 & $2.3 \times 10^{23}$ \\
\\
2015 October~27 (post-periastron)
    & 54 &  3.6 &  7.1 & 2.5 & $2.6 \times 10^{23}$\\
    & 42 &  6.8 & 13.2 & 4.3 & $3.7 \times 10^{23}$\\ 
2015 October~30 (post-periastron)
    & 54 &  1.6 &  1.9 & 1.3 & $1.4 \times 10^{23}$\\ 
    & 42 &  2.6 &  3.3 & 2.1 & $1.8 \times 10^{23}$\\
\hline
\end{tabular}
\begin{tablenotes}
 \item[a] Because of ionospheric conditions, no good data were obtained on~2015 October~24.
\end{tablenotes}
\end{threeparttable}

\caption{Image sensitivities and inferred planetary radio luminosity limits. Values for the full band (30~MHz--78~MHz) are referenced to a center frequency of~54~MHz; values for the lower half of the band (30~MHz--54~MHz) are referenced to a center frequency of~42~MHz.
Luminosity limits are obtained from the Stokes~V images.  In converting from flux density to luminosity, isotropic radiation and an emission bandwidth on the order of the observing frequency are assumed.}\label{tab:limits}
\end{table*}

Due to the reduced bandwidth, we expected the image noise levels for the lower half of the band to be approximately 40\% higher than those across the full band.  In practice, we find that the noise levels in the lower half of the band are closer to 60\% higher, a difference that we attribute to the lower sensitivity of the LBA dipoles in that frequency range \citep{VanHaarlem2013} combined with an increased level in RFI at lower frequencies and residual ionospheric calibration uncertainties producing larger effects at lower frequencies.

\begin{figure*}
  \centering
  \includegraphics[width=\textwidth]{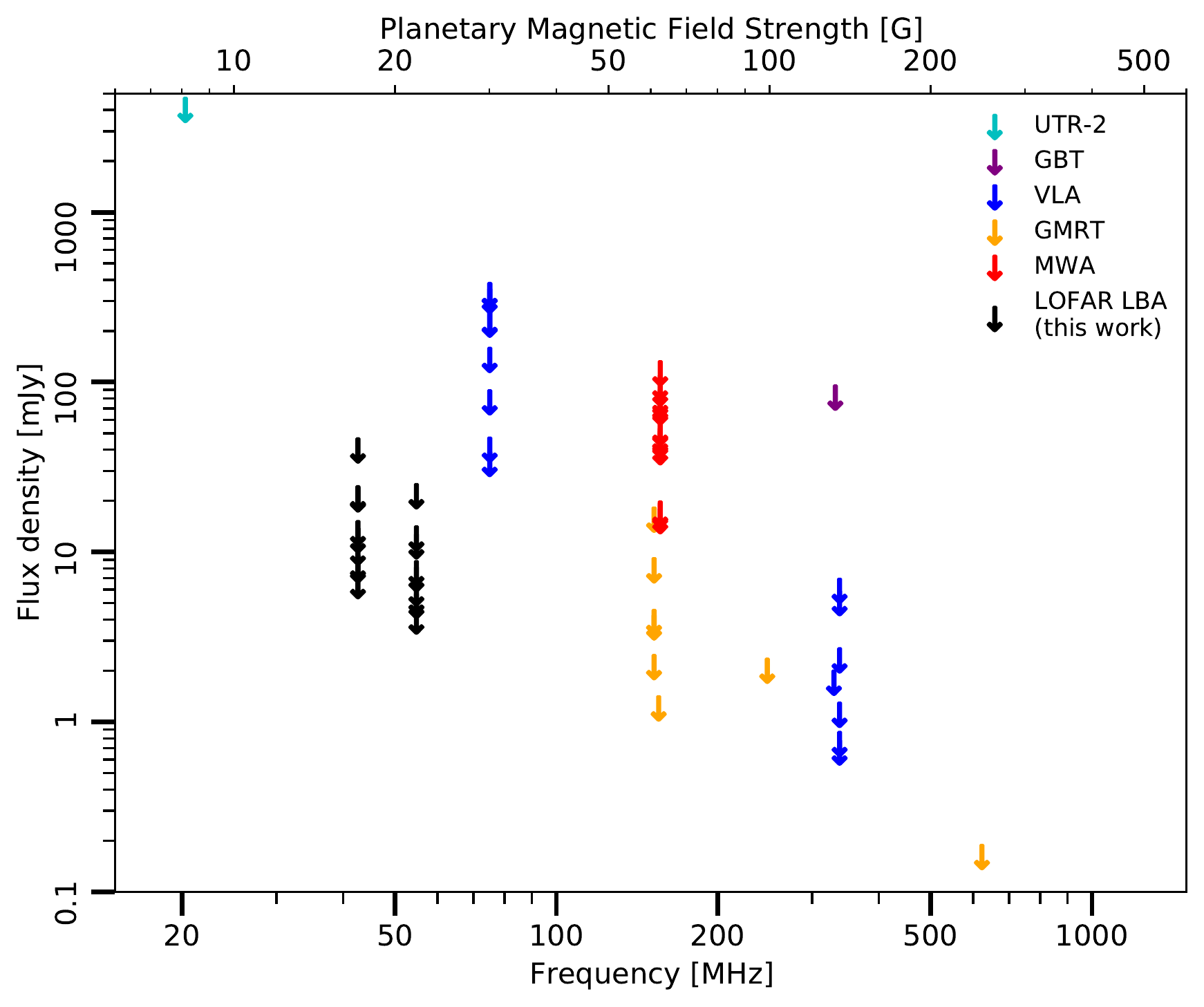}\hfil%
  \caption{Upper limits for direct observation of radio emission from extrasolar planets. In black, we show the results presented in this paper at a $3\sigma$ confidence level. For reference, at the distance of \planet{}, the burst emission of Jupiter would be weaker than 0.1~mJy.}
  \label{fig:limits}
\end{figure*}

\section{Discussion and conclusions}
\label{sec:conclude}

The results summarized in Table~\ref{tab:limits} are distinguished from previous observations of \planet{} specifically and {most} other radio observations of extrasolar planets generally by two factors: (i)~They are at frequencies near~50~MHz, much closer to the frequencies at which Jupiter emits ($\nu < 40\,\mathrm{MHz}$) and, {with the exception of observations by the Clark Lake Radio Observatory \citep{Yantis1977} and the Ukrainian T-shaped Radio telescope \citep[UTR-2,][]{2004P&SS...52.1479R},} lower than most previously reported observations of extrasolar planets; and~(ii)~sensitivities of approximately a few millijanskys have been achieved, an order of magnitude or more below nearly all previous extrasolar planet observations below~100~MHz.

\cite{Lazio2010} estimated that the radio luminosity of \planet{} could be as much as a factor of~3000 higher due to its much smaller distance to its host star as compared to Jupiter. The peak radio luminosity of Jupiter can reach $2 \times 10^{18}$~erg~s${}^{-1}$ near~40~MHz {\citep{2004JGRA..109.9S15Z}}; therefore, the radio luminosity of \planet{} might be as much as $6 \times 10^{21}$~erg~s${}^{-1}$.

Implicit in the original analysis by \cite{Lazio2010} and in our discussion heretofore has been that the interaction between the \hdstar{} stellar wind and the (assumed) magnetosphere of \planet{} is super-Alfv{\'e}nic, akin to the solar wind-planetary magnetosphere interaction for all the giant planets in the Solar System.  If the interaction is sub-Alfv{\'e}nic, the planet's magnetosphere does not sustain a bow shock. In a sub-Alfv{\'e}nic interaction, energy flows toward both the star and the planet are expected \citep[e.g.,][]{2004ApJ...602L..53I,pkbm05,pkbm06,ww05}, but whether an electron cyclotron maser instability could be sustained in the planet's polar regions is unclear. This scenario has been discussed at length, considering both Solar System bodies and extrasolar planets \citep{2001Ap&SS.277..293Z,2006A&A...460..317P,2007P&SS...55..598Z,sgdns13}. Specifically, \cite{sgdns13} appear to have considered the conditions for \planet{} when it was near apastron instead of the conditions appropriate for these observations, namely near periastron.

We now assess whether the \hdstar{} stellar wind (assumed) magnetosphere of \planet{} interaction is likely to be super-Alfv{\'e}nic or not. From the NASA Exoplanet Archive, the properties of \hdstar{} are approximately solar, with $M = 1.018\,\Msun$, $R = 1.037\,\Rsun$, and age $\approx 5.9$~Gyr \citep{2017A&A...602A.107B}, and, given the various uncertainties, we henceforth treat it as equivalent to the Sun. At the time of the observations, the planet was at a distance of about~6.5 stellar radii, equivalent to approximately 0.03~au \citep{2007P&SS...55..598Z}.
At this distance, the solar wind has a plasma mass density of about $2 \times 10^{-20}$~g~cm$^3$ \citep{1996SoPh..169..209K}, the magnetic field is about $10^{-2}$~\hbox{G}, and the solar wind velocity is about 100~\kms. The equivalent Alfv{\'e}n velocity is about 220~\kms.

The orbital velocity of \planet{} at this phase in its orbit is approximately 250~\kms. In contrast to hot Jupiters, which have essentially circular orbits at similar distances from their host stars, \planet{} will have a nearly radial velocity. Thus, the equivalent velocity at which the stellar wind impacts the magnetosphere is approximately 350~\kms. We conclude that the stellar wind-planet (magnetosphere) interaction is likely to be (moderately) super-Alfv{\'e}nic and that an electron cyclotron maser instability could be sustained at least pre-periastron.  Post-periastron, as the planet recedes from the star, there may be intervals during which the interaction becomes sub-Alfv{\'e}nic.

By conducting observations in the 30~MHz--78~MHz band, this work has addressed one of the major uncertainties present in most previous efforts to detect the radio emission from extrasolar planets. While there have been a number of relatively sensitive searches conducted, these have typically been at frequencies above~100~MHz, in comparison to Jupiter's upper frequency for emission of approximately 35~MHz. Furthermore, as discussed in \cite{Lazio2010}, estimates of the upper emission frequency for \planet{} are between 55~MHz and~90~MHz, based on determinations of various planetary parameters and scaling laws.

Broadly speaking, many of the considerations for the observations of \planet{} that were described by \cite{Lazio2010} remain valid today. \planet{} remains the planet with the second-highest known orbital eccentricity. While \object{HD~7449b} has since been discovered \citep[{$e = 0.92$,}][]{Dumusque2011, Wittenmyer2019}, of the planets with the highest orbital eccentricities, \planet{} remains the one with the shortest orbital period (thereby facilitating more opportunities for observation), and none of those planets have distances significantly shorter than \planet{}{;} \object{HD~20782b} ($e = 0.95$) is at a distance of~36~pc, resulting in less than a factor of~four improvement in potential radio luminosity limits relative to \planet.  \object{Fomalhaut~b} has a significant orbital eccentricity ($e = 0.87$) and has the virtue of being significantly closer ($\approx 7.7$~pc), but its notional orbital period is so long ($\approx 1500$~yr) and its semi-major axis so large (160~au) that it {is} not a practical target. Furthermore, recent observations have questioned its nature \citep{Gaspar9712}.
Similarly, no revolutionary improvements in observational capability seem to be coming in the near term, with the exception of the completion of {NenuFAR} \citep{z+20} and the LOFAR upgrade (LOFAR~2.0). LOFAR 2.0, planned to be operational by 2024, aims to improve the sensitivity of the LBA system, doubling the usable dipoles of LOFAR at frequencies below 100~MHz. Another improvement will come with the simultaneous observations with the LOFAR HBA system, which will be crucial for modeling ionospheric disturbances and removing their effect from LBA data. Unfortunately, the Sun is moving from the solar cycle minimum to maximum, which will likely lead to an interval of higher ionospheric activity that will make ultra-low frequency observations more challenging. When the LOFAR~2.0 system is fully operational, these limits may be able to be improved by as as much as a factor of~five.

The low radio frequency component of the Square Kilometre Array (SKA1-Low) will likely not be fully operational until the end of the 2020s, and its site in Western Australia means that it will not be able to observe \planet, though it could observe many of the other highest-eccentricity planets. Near-term space-based radio telescopes, such as the Sun Radio Interferometer Space Experiment \citep[\hbox{SunRISE};][]{Kasper2019}, will likely not have the sensitivity to detect extrasolar planetary radio emission, unless it becomes significantly stronger at frequencies below~20~MHz; if future space-based radio telescopes are larger, they may be sensitive enough to detect such emissions.

\begin{acknowledgements}
We thank A.~Vidotto for helpful discussions about solar wind models, D.~Winterhalter and W.~Farrell for the support in writing the initial LOFAR proposal, and the referee for several comments that clarified both the content and presentation of this work.

This work is partly funded by the Deutsche Forschungsgemeinschaft under Germany's Excellence Strategy EXC~2121 ``Quantum Universe'' 390833306. Part of this research was carried out at the Jet Propulsion Laboratory, California Institute of Technology, under a contract with the National Aeronautics and Space Administration.
A portion of this work was funded by the NASA Lunar Science Institute (via Cooperative Agreement NNA09DB30A).

This research has made use of the NASA Exoplanet Archive, which is operated by the California Institute of Technology, under contract with the National Aeronautics and Space Administration under the Exoplanet Exploration Program.  This research has made use of NASA’s Astrophysics Data System Bibliographic Services.

LOFAR is the LOw Frequency ARray designed and constructed by \hbox{ASTRON}. It has observing, data processing, and data storage facilities in several countries, which are owned by various parties (each with their own funding sources), and are collectively operated by the ILT foundation under a joint scientific policy. The ILT resources have benefitted from the following recent major funding sources: \hbox{CNRS-INSU}, Observatoire de Paris and Universite d’Orleans, France; \hbox{BMBF}, \hbox{MIWF-NRW}, \hbox{MPG}, Germany; Science Foundation Ireland
(SFI), Department of Business, Enterprise and Innovation
(DBEI), Ireland; \hbox{NWO}, The Netherlands; The Science and
Technology Facilities Council, \hbox{UK}; Ministry of Science and Higher Education, Poland; Istituto Nazionale di Astrofisica (INAF).

\end{acknowledgements}

\bibliographystyle{aa}
\bibliography{library}

\begin{thebibliography}{51}
\expandafter\ifx\csname natexlab\endcsname\relax\def\natexlab#1{#1}\fi

\bibitem[{{Bastian} {et~al.}(2000){Bastian}, {Dulk}, \&
  {Leblanc}}]{2000ApJ...545.1058B}
{Bastian}, T.~S., {Dulk}, G.~A., \& {Leblanc}, Y. 2000, \apj, 545, 1058

\bibitem[{{Bonomo} {et~al.}(2017){Bonomo}, {Desidera}, {Benatti}, {Borsa},
  {Crespi}, {Damasso}, {Lanza}, {Sozzetti}, {Lodato}, {Marzari}, {Boccato},
  {Claudi}, {Cosentino}, {Covino}, {Gratton}, {Maggio}, {Micela}, {Molinari},
  {Pagano}, {Piotto}, {Poretti}, {Smareglia}, {Affer}, {Biazzo}, {Bignamini},
  {Esposito}, {Giacobbe}, {H{\'e}brard}, {Malavolta}, {Maldonado}, {Mancini},
  {Martinez Fiorenzano}, {Masiero}, {Nascimbeni}, {Pedani}, {Rainer}, \& {Scand
  ariato}}]{2017A&A...602A.107B}
{Bonomo}, A.~S., {Desidera}, S., {Benatti}, S., {et~al.} 2017, \aap, 602, A107

\bibitem[{{Cauley} {et~al.}(2019){Cauley}, {Shkolnik}, {Llama}, \&
  {Lanza}}]{2019NatAs...3.1128C}
{Cauley}, P.~W., {Shkolnik}, E.~L., {Llama}, J., \& {Lanza}, A.~F. 2019, Nature
  Astronomy, 3, 1128

\bibitem[{Condon {et~al.}(1998)Condon, Cotton, Greisen, Yin, Perley, Taylor, \&
  Broderick}]{Condon1998}
Condon, J.~J., Cotton, W.~D., Greisen, E.~W., {et~al.} 1998, Astron. J., 8065,
  1693

\bibitem[{{de Gasperin} {et~al.}(2020){de Gasperin}, {Brunetti}, {Br{\"u}ggen},
  {van Weeren}, {Williams}, {Botteon}, {Cuciti}, {Dijkema}, {Edler},
  {Iacobelli}, {Kang}, {Offringa}, {Orr{\'u}}, {Pizzo}, {Rafferty},
  {R{\"o}ttgering}, \& {Shimwell}}]{2020A&A...642A..85D}
{de Gasperin}, F., {Brunetti}, G., {Br{\"u}ggen}, M., {et~al.} 2020, \aap, 642,
  A85

\bibitem[{de~Gasperin {et~al.}(2019)de~Gasperin, Dijkema, Drabent, Mevius,
  Rafferty, van Weeren, Br{\"{u}}ggen, Callingham, Emig, Heald, Intema,
  Morabito, Offringa, Oonk, Orr{\`{u}}, R{\"{o}}ttgering, Sabater, Shimwell,
  Shulevski, \& Williams}]{deGasperin2019}
de~Gasperin, F., Dijkema, T.~J., Drabent, A., {et~al.} 2019, A{\&}A, 5, A5

\bibitem[{{Desch} \& {Barrow}(1984)}]{1984JGR....89.6819D}
{Desch}, M.~D. \& {Barrow}, C.~H. 1984, \jgr, 89, 6819

\bibitem[{{Desch} \& {Rucker}(1983)}]{1983JGR....88.8999D}
{Desch}, M.~D. \& {Rucker}, H.~O. 1983, \jgr, 88, 8999

\bibitem[{Dumusque {et~al.}(2011)Dumusque, Lovis, S{\'{e}}gransan, Mayor, Udry,
  Benz, Bouchy, {Lo Curto}, Mordasini, Pepe, Queloz, Santos, \&
  Naef}]{Dumusque2011}
Dumusque, X., Lovis, C., S{\'{e}}gransan, D., {et~al.} 2011, A{\&}A, 535

\bibitem[{{Ekenb{\"a}ck} {et~al.}(2010){Ekenb{\"a}ck}, {Holmstr{\"o}m}, {Wurz},
  {Grie{\ss}meier}, {Lammer}, {Selsis}, \& {Penz}}]{2010ApJ...709..670E}
{Ekenb{\"a}ck}, A., {Holmstr{\"o}m}, M., {Wurz}, P., {et~al.} 2010, \apj, 709,
  670

\bibitem[{{Farrell} {et~al.}(1999){Farrell}, {Desch}, \&
  {Zarka}}]{1999JGR...10414025F}
{Farrell}, W.~M., {Desch}, M.~D., \& {Zarka}, P. 1999, \jgr, 104, 14025

\bibitem[{{Fossey} {et~al.}(2009){Fossey}, {Waldmann}, \&
  {Kipping}}]{2009MNRAS.396L..16F}
{Fossey}, S.~J., {Waldmann}, I.~P., \& {Kipping}, D.~M. 2009, \mnras, 396, L16

\bibitem[{{France} {et~al.}(2010){France}, {Stocke}, {Yang}, {Linsky},
  {Wolven}, {Froning}, {Green}, \& {Osterman}}]{2010ApJ...712.1277F}
{France}, K., {Stocke}, J.~T., {Yang}, H., {et~al.} 2010, \apj, 712, 1277

\bibitem[{{Gallagher} \& {Dangelo}(1981)}]{1981GeoRL...8.1087G}
{Gallagher}, D.~L. \& {Dangelo}, N. 1981, \grl, 8, 1087

\bibitem[{G{\'a}sp{\'a}r \& Rieke(2020)}]{Gaspar9712}
G{\'a}sp{\'a}r, A. \& Rieke, G.~H. 2020, Proceedings of the National Academy of
  Sciences, 117, 9712

\bibitem[{{Hut}(1981)}]{1981A&A....99..126H}
{Hut}, P. 1981, \aap, 99, 126

\bibitem[{Intema {et~al.}(2017)Intema, Jagannathan, Mooley, \&
  Frail}]{Intema2017}
Intema, H.~T., Jagannathan, P., Mooley, K.~P., \& Frail, D.~A. 2017, A{\&}A,
  598, A78

\bibitem[{{Ip} {et~al.}(2004){Ip}, {Kopp}, \& {Hu}}]{2004ApJ...602L..53I}
{Ip}, W.-H., {Kopp}, A., \& {Hu}, J.-H. 2004, \apjl, 602, L53

\bibitem[{Kao {et~al.}(2016)Kao, Hallinan, Pineda, Escala, Burgasser, Bourke,
  \& Stevenson}]{Kao_2016}
Kao, M.~M., Hallinan, G., Pineda, J.~S., {et~al.} 2016, Astrophys. J., 818, 24

\bibitem[{Kasper {et~al.}(2019)Kasper, Lazio, Romero-Wolf, Lux, \&
  Neilsen}]{Kasper2019}
Kasper, J., Lazio, J., Romero-Wolf, A., Lux, J., \& Neilsen, T. 2019, in IEEE
  Aerosp. Conf. Proc., Vol. 2019-March (IEEE Computer Society)

\bibitem[{{Knapp}(2018)}]{k18}
{Knapp}, M.~E. 2018, PhD thesis, Massachusetts Institute of Technology

\bibitem[{{K{\"o}hnlein}(1996)}]{1996SoPh..169..209K}
{K{\"o}hnlein}, W. 1996, \solphys, 169, 209

\bibitem[{Lane {et~al.}(2014)Lane, Cotton, van Velzen, Clarke, Kassim,
  Helmboldt, Lazio, \& Cohen}]{Lane2014}
Lane, W.~M., Cotton, W.~D., van Velzen, S., {et~al.} 2014, MNRAS, 440, 327

\bibitem[{{Lazio} \& {Farrell}(2007)}]{2007ApJ...668.1182L}
{Lazio}, T. J.~W. \& {Farrell}, W.~M. 2007, \apj, 668, 1182

\bibitem[{Lazio {et~al.}(2010)Lazio, Shankland, Farrell, \& Blank}]{Lazio2010}
Lazio, T. J.~W., Shankland, P.~D., Farrell, W.~M., \& Blank, D.~L. 2010,
  Astron. J., 140, 1929

\bibitem[{{Lazio} {et~al.}(2016){Lazio}, {Shkolnik}, {Hallinan}, \& {Planetary
  Habitability Study Team}}]{2016pmf..rept.....L}
{Lazio}, T.~J.~W., {Shkolnik}, E., {Hallinan}, G., \& {Planetary Habitability
  Study Team}. 2016, {Planetary Magnetic Fields: Planetary Interiors and
  Habitability}, W.~M.~Keck Institute for Space Studies: Planetary Magnetic
  Fields: Planetary Interiors and Habitability

\bibitem[{Offringa {et~al.}(2014)Offringa, McKinley, Hurley-Walker, Briggs,
  Wayth, Kaplan, Bell, Feng, Neben, Hughes, Rhee, Murphy, Bhat, Bernardi,
  Bowman, Cappallo, Corey, Deshpande, Emrich, Ewall-Wice, Gaensler, Goeke,
  Greenhill, Hazelton, Hindson, Johnston-Hollitt, Jacobs, Kasper, Kratzenberg,
  Lenc, Lonsdale, Lynch, McWhirter, Mitchell, Morales, Morgan, Kudryavtseva,
  Oberoi, Ord, Pindor, Procopio, Prabu, Riding, Roshi, Shankar, Srivani,
  Subrahmanyan, Tingay, Waterson, Webster, Whitney, Williams, \&
  Williams}]{Offringa2014}
Offringa, A.~R., McKinley, B., Hurley-Walker, N., {et~al.} 2014, MNRAS, 444,
  606

\bibitem[{{Preusse} {et~al.}(2005){Preusse}, {Kopp}, {B\"uchner}, \&
  {Motschmann}}]{pkbm05}
{Preusse}, S., {Kopp}, A., {B\"uchner}, J., \& {Motschmann}, U. 2005, A\&A,
  434, 1191

\bibitem[{{Preusse} {et~al.}(2006{\natexlab{a}}){Preusse}, {Kopp},
  {B{\"u}chner}, \& {Motschmann}}]{2006A&A...460..317P}
{Preusse}, S., {Kopp}, A., {B{\"u}chner}, J., \& {Motschmann}, U.
  2006{\natexlab{a}}, \aap, 460, 317

\bibitem[{{Preusse} {et~al.}(2006{\natexlab{b}}){Preusse}, {Kopp}, {B\"uchner},
  \& {Motschmann}}]{pkbm06}
{Preusse}, S., {Kopp}, A., {B\"uchner}, J., \& {Motschmann}, U.
  2006{\natexlab{b}}, A\&A, 460, 317

\bibitem[{Rengelink {et~al.}(1997)Rengelink, Tang, de~Bruyn, Miley, a.R.
  Bremer, Rottgering, a.R. Bremer, R�ttgering, \& a.R.
  Bremer}]{Rengelink1997}
Rengelink, R.~B., Tang, Y., de~Bruyn, a.~G., {et~al.} 1997, Astron. Astrophys.
  Suppl., 124, 259

\bibitem[{{Ryabov} {et~al.}(2004){Ryabov}, {Zarka}, \&
  {Ryabov}}]{2004P&SS...52.1479R}
{Ryabov}, V.~B., {Zarka}, P., \& {Ryabov}, B.~P. 2004, \planss, 52, 1479

\bibitem[{{Saur} {et~al.}(2013){Saur}, {Grambusch}, {Duling}, {Neubauer}, \&
  {Simon}}]{sgdns13}
{Saur}, J., {Grambusch}, T., {Duling}, S., {Neubauer}, F.~M., \& {Simon}, S.
  2013, A\&A, 552, A119

\bibitem[{{Treumann}(2006)}]{2006A&ARv..13..229T}
{Treumann}, R.~A. 2006, \aapr, 13, 229

\bibitem[{van Haarlem {et~al.}(2013)van Haarlem, Wise, Gunst, Heald, McKean,
  Hessels, de~Bruyn, Nijboer, Swinbank, Fallows, Brentjens, Nelles, Beck,
  Falcke, Fender, H{\"{o}}randel, Koopmans, Mann, Miley, R{\"{o}}ttgering,
  Stappers, Wijers, Zaroubi, van~den Akker, Alexov, Anderson, Anderson, van
  Ardenne, Arts, Asgekar, Avruch, Batejat, B{\"{a}}hren, Bell, Bell, van
  Bemmel, Bennema, Bentum, Bernardi, Best, B{\^{i}}rzan, Bonafede,
  a.~J.~Boonstra, Braun, Bregman, Breitling, van~de Brink, Broderick, Broekema,
  Brouw, Br{\"{u}}ggen, Butcher, van Cappellen, Ciardi, Coenen, Conway, Coolen,
  Corstanje, Damstra, Davies, Deller, Dettmar, van Diepen, Dijkstra, Donker,
  Doorduin, Dromer, Drost, van Duin, Eisl{\"{o}}ffel, van Enst, Ferrari,
  Frieswijk, Gankema, Garrett, de~Gasperin, Gerbers, de~Geus, Grie{\ss}meier,
  Grit, Gruppen, Hamaker, Hassall, Hoeft, Holties, Horneffer, van~der Horst,
  van Houwelingen, Huijgen, Iacobelli, Intema, Jackson, Jelic, de~Jong, Juette,
  Kant, Karastergiou, Koers, Kollen, Kondratiev, Kooistra, Koopman, Koster,
  Kuniyoshi, Kramer, Kuper, Lambropoulos, Law, van Leeuwen, Lemaitre, Loose,
  Maat, Macario, Markoff, Masters, McFadden, McKay-Bukowski, Meijering,
  Meulman, Mevius, Middelberg, Millenaar, Miller-Jones, Mohan, Mol, Morawietz,
  Morganti, Mulcahy, Mulder, Munk, Nieuwenhuis, van Nieuwpoort, Noordam,
  Norden, Noutsos, Offringa, Olofsson, Omar, Orr{\'{u}}, Overeem, Paas,
  Pandey-Pommier, Pandey, Pizzo, Polatidis, Rafferty, Rawlings, Reich,
  de~Reijer, Reitsma, Renting, Riemers, Rol, Romein, Roosjen, Ruiter, Scaife,
  van~der Schaaf, Scheers, Schellart, Schoenmakers, Schoonderbeek, Serylak,
  Shulevski, Sluman, Smirnov, Sobey, Spreeuw, Steinmetz, Sterks, Stiepel,
  Stuurwold, Tagger, Tang, Tasse, Thomas, Thoudam, Toribio, van~der Tol, Usov,
  van Veelen, van~der Veen, ter Veen, Verbiest, Vermeulen, Vermaas, Vocks,
  Vogt, de~Vos, van~der Wal, van Weeren, Weggemans, Weltevrede, White,
  Wijnholds, Wilhelmsson, Wucknitz, Yatawatta, Zarka, Zensus, \& van
  Zwieten}]{VanHaarlem2013}
van Haarlem, M.~P., Wise, M.~W., Gunst, a.~W., {et~al.} 2013, A{\&}A, 556, A2

\bibitem[{{Vedantham} {et~al.}(2020){Vedantham}, {Callingham}, {Shimwell},
  {Tasse}, {Pope}, {Bedell}, {Snellen}, {Best}, {Hardcastle}, {Haverkorn},
  {Mechev}, {O'Sullivan}, {R{\"o}ttgering}, \& {White}}]{2020NatAs...4..577V}
{Vedantham}, H.~K., {Callingham}, J.~R., {Shimwell}, T.~W., {et~al.} 2020,
  Nature Astronomy, 4, 577

\bibitem[{{Vidotto} {et~al.}(2012){Vidotto}, {Fares}, {Jardine}, {Donati},
  {Opher}, {Moutou}, {Catala}, \& {Gombosi}}]{2012MNRAS.423.3285V}
{Vidotto}, A.~A., {Fares}, R., {Jardine}, M., {et~al.} 2012, \mnras, 423, 3285

\bibitem[{{Vorgul} {et~al.}(2011){Vorgul}, {Kellett}, {Cairns}, {Bingham},
  {Ronald}, {Speirs}, {McConville}, {Gillespie}, \&
  {Phelps}}]{2011PhPl...18e6501V}
{Vorgul}, I., {Kellett}, B.~J., {Cairns}, R.~A., {et~al.} 2011, Physics of
  Plasmas, 18, 056501

\bibitem[{{Weber} {et~al.}(2017){Weber}, {Lammer}, {Shaikhislamov}, {Chadney},
  {Khodachenko}, {Grie{\ss}meier}, {Rucker}, {Vocks}, {Macher}, {Odert}, \&
  {Kislyakova}}]{2017MNRAS.469.3505W}
{Weber}, C., {Lammer}, H., {Shaikhislamov}, I.~F., {et~al.} 2017, \mnras, 469,
  3505

\bibitem[{{Willes} \& {Wu}(2005)}]{ww05}
{Willes}, A.~J. \& {Wu}, K. 2005, A\&A, 432, 1091

\bibitem[{{Winglee} {et~al.}(1986){Winglee}, {Dulk}, \&
  {Bastian}}]{1986ApJ...309L..59W}
{Winglee}, R.~M., {Dulk}, G.~A., \& {Bastian}, T.~S. 1986, \apjl, 309, L59

\bibitem[{Wittenmyer {et~al.}(2019)Wittenmyer, Clark, Zhao, Horner, Wang, \&
  Johns}]{Wittenmyer2019}
Wittenmyer, R.~A., Clark, J.~T., Zhao, J., {et~al.} 2019, MNRAS, 484, 5859

\bibitem[{Yadav \& Thorngren(2017)}]{Yadav_2017}
Yadav, R.~K. \& Thorngren, D.~P. 2017, Astrophys. J., 849, L12

\bibitem[{Yantis {et~al.}(1977)Yantis, {Sullivan, W. T.}, \&
  Erickson}]{Yantis1977}
Yantis, W.~F., {Sullivan, W. T.}, I., \& Erickson, W.~C. 1977, Bull. Am.
  Astron. Soc. Vol. 9, p.453, 9, 453

\bibitem[{{Zarka}(2007)}]{2007P&SS...55..598Z}
{Zarka}, P. 2007, \planss, 55, 598

\bibitem[{{Zarka} {et~al.}(2004){Zarka}, {Cecconi}, \&
  {Kurth}}]{2004JGRA..109.9S15Z}
{Zarka}, P., {Cecconi}, B., \& {Kurth}, W.~S. 2004, Journal of Geophysical
  Research (Space Physics), 109, A09S15

\bibitem[{{Zarka} {et~al.}(2020){Zarka}, {Denis}, {Tagger}, {Girard}, {Coffre},
  {Dumez-Viou}, {Taffoureau}, {Charrier}, {Bondonneau}, {Briand}, {Casoli},
  {Cecconi}, {Cognard}, {Corbel}, {Dallier}, {Ferrari}, {Griessmeier}, {Loh},
  {Martin}, {Pommier}, {Semelin}, {Tasse}, {Theureau}, {Tremou}, {Hellbourg},
  {Konovalenko}, {Koopmans}, {Tokarsky}, {Ulyanov}, {Vermeulen}, {Zakharenko},
  \& {NenuFAR-France team}}]{z+20}
{Zarka}, P., {Denis}, L., {Tagger}, M., {et~al.} 2020, in International Union
  of Radio Science, General Assembly \& Science Symposium 2020, URSI GASS 2020,
  Rome, Italy,
  http://www.ursi.org/proceedings/procGA20/papers/URSIGASS2020SummaryPaperNenuFARnew.pdf

\bibitem[{{Zarka} {et~al.}(2015){Zarka}, {Lazio}, \&
  {Hallinan}}]{2015aska.confE.120Z}
{Zarka}, P., {Lazio}, J., \& {Hallinan}, G. 2015, in Advancing Astrophysics
  with the Square Kilometre Array (AASKA14), 120

\bibitem[{{Zarka} {et~al.}(2018){Zarka}, {Marques}, {Louis}, {Ryabov}, {Lamy},
  {Echer}, \& {Cecconi}}]{2018A&A...618A..84Z}
{Zarka}, P., {Marques}, M.~S., {Louis}, C., {et~al.} 2018, \aap, 618, A84

\bibitem[{{Zarka} {et~al.}(1997){Zarka}, {Queinnec}, {Ryabov}, {Ryabov},
  {Shevchenko}, {Arkhipov}, {Rucker}, {Denis}, {Gerbault}, {Dierich}, \&
  {Rosolen}}]{1997pre4.conf..101Z}
{Zarka}, P., {Queinnec}, J., {Ryabov}, B.~P., {et~al.} 1997, in Planetary Radio
  Emission IV, 101--127

\bibitem[{{Zarka} {et~al.}(2001){Zarka}, {Treumann}, {Ryabov}, \&
  {Ryabov}}]{2001Ap&SS.277..293Z}
{Zarka}, P., {Treumann}, R.~A., {Ryabov}, B.~P., \& {Ryabov}, V.~B. 2001,
  \apss, 277, 293

\end{thebibliography}

\end{document}